\documentclass[twocolumn,english,aps,prl,showpacs,superscriptaddress,amssymb,amsfonts]{revtex4-1}
\usepackage[T1]{fontenc}
\usepackage[latin9]{inputenc}
\usepackage{bm}
\usepackage{amsmath}
\usepackage{graphicx}
\usepackage{amssymb}
\usepackage{esint}
\usepackage{units}
\usepackage{natbib}
\usepackage{hyperref}

\makeatletter
\@ifundefined{textcolor}{}
{%
 \definecolor{BLACK}{gray}{0}
 \definecolor{WHITE}{gray}{1}
 \definecolor{RED}{rgb}{1,0,0}
 \definecolor{GREEN}{rgb}{0,1,0}
 \definecolor{BLUE}{rgb}{0,0,1}
 \definecolor{CYAN}{cmyk}{1,0,0,0}
 \definecolor{MAGENTA}{cmyk}{0,1,0,0}
 \definecolor{YELLOW}{cmyk}{0,0,1,0}
 }

\newcommand{\bra}[1]{ {\left\langle {#1} \right|} }
\newcommand{\ket}[1]{ {\left| {#1} \right\rangle} }

\newcommand{\vk}{ {\boldsymbol{k}} }
\renewcommand{\vr}{ {\boldsymbol{r}} }

\makeatother

\usepackage{babel}

\begin{document}

\title{Majorana modes at the vortex ends of superconducting doped topological
insulators}

\author{Pavan Hosur}

\affiliation{Department of Physics, University of California, Berkeley, CA 94720,
USA}

\author{Pouyan Ghaemi}

\affiliation{Department of Physics, University of California, Berkeley, CA 94720,
USA}

\affiliation{Materials Sciences Division, Lawrence Berkeley National Laboratory,
Berkeley, CA 94720}

\author{Roger S.~K. Mong}

\affiliation{Department of Physics, University of California, Berkeley, CA 94720,
USA}

\author{Ashvin Vishwanath}

\affiliation{Department of Physics, University of California, Berkeley, CA 94720,
USA}

\affiliation{Materials Sciences Division, Lawrence Berkeley National Laboratory,
Berkeley, CA 94720}
\begin{abstract}

Recent experiments have observed bulk superconductivity in doped topological
insulators. Here we ask whether vortex Majorana zero modes, previously
predicted to occur when superconductivity is induced on the surface
of topological insulators, survive even in these doped systems with
metallic normal states. Assuming inversion symmetry, we find that Majorana zero modes indeed appear but only below
a critical doping. The critical doping is associated with a topological phase transition
of the vortex line, where it supports gapless excitations along its
length.
%For weak pairing, the critical point in a simple model is
%shown to occur at the doping level where band inversion reverses sign.
%More generally, it
The critical point depends only on the orientation of the
vortex line, and a Berry phase property, the SU(2) Berry phase of the Fermi
surface in the metallic normal state. By calculating this phase for available band structures we determine that materials candidates like
$p$-doped Bi$_{2}$Te$_{3}$ under pressure supports vortex end Majorana
modes. Surprisingly, even superconductors derived from topologically trivial band structures can support Majorana modes, providing a promising route to realizing them.
\end{abstract}
\maketitle

Majorana fermions, defined as fermions that are their own anti-particles
unlike conventional Dirac fermions such as electrons, have long been
sought by high energy physicists, but so far in vain. Of late, the search for Majorana fermions has remarkably shifted to condensed matter systems~\cite{KitaevWireMajorana01,dassarma,ReadGreenP+ipFQHE00}, especially, to superconductors (SCs), where states appear in conjugate pairs with equal and opposite energies. Then, a single state at zero energy is its own conjugate and hence, a Majorana state or a \emph{Majorana zero mode} ({\bf MZM}).
%As the particle-hole symmetry in a SC cannot be broken by a local disturbance, t
These states are immune to local noise and hence, considered strong candidates for storing quantum information and performing fault tolerant quantum computation~\cite{NayakNonAbelianQuantComp}. Moreover, they show non-Abelian rather than Bose or Fermi statistics which leads to a number of extraordinary phenomenon~\cite{MooreReadNonabelion91}.

Despite many proposals %~\cite{KitaevWireMajorana01,dassarma},
direct experimental evidence for a MZM is still lacking. While initial proposals involved the
$\nu=5/2$ quantum Hall state and SCs with unconventional
pairing such as $p_{x}+ip_{y}$~\cite{ReadGreenP+ipFQHE00}, a recent breakthrough occurred with
the discovery of topological insulators (TIs)~\cite{HasanKaneReview,*QiZhangReview,*HasanMooreReview},
which feature metallic surface bands. When a conventional s-wave
SC is brought near this metallic surface, a single MZM
is trapped in the vortex core~\cite{FuKaneSCProximity}. Since then,
several TIs were found to exhibit {\em bulk} superconductivity
on doping~\cite{SCdoping} or under pressure~\cite{pressuresc,ZhangBi2Te3SCunderPressure}.
The normal phase of these SCs is now metallic, which raises the question:
can a SC vortex host a surface MZM even when the bulk
is {\em not} insulating?

In this Letter, we answer this question in the affirmative and in the process, discover a convenient way to obtain a MZM, which allows us to conclude that some existing experimental systems should possess these states. Our proposal involves simply passing a magnetic field through a TI-based SC, such as superconducting Bi$_2$Te$_3$, in which the doping is below a certain threshold value. We also find general criteria for SCs to host vortex MZMs and show that some non-TI-based SCs satisfy them too.

A heuristic rule often applied to answer the above question is to examine
whether the normal state bulk Fermi surface (FS) is well separated
from surface states in the Brillouin zone. If it is, MZMs are
assumed to persist in the bulk SC. While this may indicate the presence
of low energy states, it is not a topological criterion since it depends
on non-universal details of surface band structure, and cannot signal
the presence of true MZMs. For MZMs to disappear, a gapless channel
must open that allows pairs to approach each other and annihilate.
%Perturbations that only modify surface state properties should not
%affect MZMs as long as the gap remains open. 
We therefore search for
and offer a bulk rather than surface criterion. In this process, we
have uncovered the following interesting facts. We assume inversion ($\mathcal I$)
and time reversal ($\mathcal T$) symmetric band structures, and weak pairing, since
these lead to a technical simplification and capture many real
systems. {\em (i)} The appearance of surface MZMs is tied to the
topological state of the vortex, viewed as a 1D topological SC. The critical
point at which they disappear is linked to a vortex phase transition
\textbf{(VPT)} where this topology changes. If verified, this may
be the first instance of a phase transition inside a topological defect.
{\em (ii)} The topological state of a vortex depends in general
on its orientation. {\em (iii)} Symmetry dictates that the normal state FS is doubly degenerate,
leading to an $SU(2)$ non-Abelian Berry phase~\cite{WilczekZee}
for closed curves, which determines the condition for quantum criticality of the vortex. This is a rare example of a {\em{non-Abelian}} Berry phase directly influencing measurable physical properties of an electronic system. Spin-orbit coupling is essential to obtaining the SU(2) Berry Phase. {\em (iv)} Using
this criterion and available band structures we find that MZMs occur in $p$-doped superconducting Bi$_{2}$Te$_{3}$~\cite{ZhangBi2Te3SCunderPressure} and in Cu-doped Bi$_{2}$Se$_{3}$ \cite{SCdoping} if the vortex is sufficiently tilted off the c-axis. C-axis vortices in Cu-doped Bi$_{2}$Se$_{3}$ are predicted to be near the topological transition.

%A key observation is that the stability of the surface MZMs is tied to the topological phase of the vortex, viewed as a 1D system. The critical point at which they disappear is linked to a vortex phase transition {\bf (VPT)} where this topology changes. We specialize to the case of inversion and time reversal symmetric band structures, since these lead to a technical simplification and capture many experimental systems.
%In the limit of weak pairing, the transition point is determined from a Berry phase criterion of a section of the normal state FS, which is selected by vortex orientation. Using this criterion and available band structures we find that c-axis vortices in Cu doped Bi$_2$Se$_3$ \cite{SCdoping} are likely to be topologically trivial, but those in $p$-doped Bi$_2$Te$_3$ which become superconducting under pressure~\cite{ZhangBi2Te3SCunderPressure}, have vortices with Majorana ends states.
%We assume  `$s$-wave' even parity pairing  and neglect the effect of magnetic field used to generate the vortices, extreme type II limit.
%In future work we will relax this and the symmetry assumptions.

{\em The Problem:} Consider a 3D insulating band structure $H$,
which we dope by changing the chemical potential $\mu$ away from
the middle of the band gap. Now, add conventional
`$s$-wave' even parity pairing $\Delta$ (in contrast to the
$p$-wave pairing of Ref. \cite{FuBergTopoSC10}) %~\footnote{Note, we do not consider topological superconductors here, which require
%a nontrivial pairing structure, such as odd parity pairing~\cite{FuBergTopoSC10}.%
%}.
%via the mean field Hamiltonian:
%\begin{equation}
%H = \sum_{rr',\,\sigma\sigma'}c^\dagger_{r\sigma}(H^I_{r\sigma,\,r'\sigma'}-\mu)c_{r'\sigma'} +\left [ \Delta_r c^\dagger_{r\sigma}\epsilon_{\sigma\sigma'}c^\dagger_{r'\sigma'}+{\rm h.c.}\right ]
%\end{equation}
%where ${\epsilon}$ is the 2x2 antisymmetric matrix.
and introduce a single vortex line into the pairing function $\Delta(\vr)$,
stretching between the top and bottom surfaces. We neglect the effect
of the magnetic field used to generate the vortices, assuming extreme
type II limit. When $H$ is a strong TI, and $\mu$ is in the band
gap, the pair potential primarily induces superconductivity on the
surface states. In this limit it is known~\cite{FuKaneSCProximity}
that MZMs appear on the surface, in the vortex
core. %
\footnote{It is convenient to discuss pairing over the entire range of $\mu$,
using the mean field Hamiltonian \eqref{eq:MeanFieldHam}, although
in reality superconductivity only appears once bulk carriers are induced.} 
Now consider tuning $\mu$ deep into the bulk bands.
By modifying states well below $\mu$, one could tune the
band structure to one with uninverted bands. One now expects `normal'
behavior, and the absence of MZMs. Therefore, a quantum
phase transition must occur between these limits and $\mu=\mu_c$.

To understand the nature of the transition, we recall some basic
facts of vortex electronic structure, which are also derived below.
Once $\mu$ enters the bulk bands, low energy Caroli-de~Gennes-Matricon
excitations appear, bound to the vortex line~\cite{CdGMSCVortexStates64}.
These excitations are still typically gapped, although by
a small energy scale, the `mini-gap': $\delta\sim\Delta/(k_{F}\xi)$,
(where $k_{F}$ is the Fermi wave-vector and $\xi$ is the coherence
length. In the weak pairing limit $k_{F}\xi\gg1$). This small energy
scale arises because the gap vanishes in the vortex core leading to
a droplet of normal fluid, which is eventually gapped by the finite
vortex size. However, the presence of the minigap is important, since
it blocks the tunneling of the surface MZMs into the vortex line,
and confines them near the surface. The closing of the mini-gap
allows the surface MZMs to tunnel along the vortex line annihilate.

{\em Vortex as a 1D topological SC:}
The VPT may be viewed as a change in the topology of the electronic
structure of the vortex line. The relevant energy scale is of the
order of the mini-gap $\delta\ll\Delta$, with excitations localized
within the 1D vortex core. The vortex admits particle-hole symmetry
($\mathcal{C}$) but breaks $\mathcal{T}$-symmetry and hence,
belongs to class D of the Altland-Zirnbauer classification~\cite{AltlandZirnbauer97}.
Thus, the problem reduces to classifying gapped phases in 1D within
the symmetry class D, which are known to be distinguished by a $\mathbb{Z}_{2}$
topological invariant~\cite{KitaevWireMajorana01}. %The symmetries of the problem allow us to view this as a superconductor with no spin rotation or time reversal symmetry (which is broken by the presence of the vortex), which is termed class D~\cite{AltlandZirnbauer97}.
%Moreover, if we are interested in low energy excitations within the pairing gap ($E<\Delta_{0}$, the pairing gap scale far away from the vortex), these are confined close to the vortex line, defining a one dimensional system.
%At the lowest energy scales we expect typically a vortex mini-gap, as discussed above.
%Thus, the problem reduces to classifying the different gapped phases of a one dimensional object - the vortex line - in symmetry class D.
%It is well known that there is a $\mathbb{Z}_{2}$ classification of such states, as pointed out by Kitaev~\cite{KitaevWireMajorana01}.
%}
The two kinds of phases differ in whether they support MZMs at their
ends. The topologically nontrivial phase does and hence, corresponds to the $\mu<\mu_{c}$
phase of the vortex line.
On raising $\mu$, the vortex line transitions into
the trivial phase, via a quantum critical point at which it is gapless
along its length. This is reminiscent of recent proposals to generate
MZMs at the ends of superconducting quantum wires \cite{DasSarmaMajoranaSCSCJunction,*Gil,*Beenakker,*Lee}.
Note, since there is no `local' gap in the vortex core, the powerful
defect topology classification of \cite{TeoKaneTopologicalDefects}
cannot be applied.

{\em The Hamiltonian} is $\mathbf{H}=\frac{1}{2}\sum_{\vk}\Psi_{\vk}^{\dagger}\mathcal{H}_{\vk}^{\textrm{BdG}}\Psi_{\vk}$
where $\Psi_{\vk}^{\dagger}=({\rm {\bf c_{\vk\uparrow}^{\dagger},\, c_{\vk\downarrow}^{\dagger},\, c_{-\vk\downarrow},\,-c_{-\vk\uparrow}}})$
and ${\rm {\bf c_{\vk\sigma}^{\dagger}}}$ is assumed to have $a=1\dots N$
orbital components ${\rm c}_{\vk\sigma a}^{\dagger}$ and \begin{align}
\mathcal{H}_{\vk}^{\textrm{BdG}}=\begin{bmatrix}H_{\vk}-\mu & \Delta\\
\Delta^{\ast} & \mu-H_{\vk}\end{bmatrix}.\label{eq:MeanFieldHam}\end{align}
 where scalars like $\mu$ and
$\Delta$ multiply the identity matrix ${\textbf{1}}_{2N\times2N}$.
The band Hamiltonian, $H_{\vk}$, is a $2N\times2N$ matrix with ${\mathcal{T}}$
symmetry: $\sigma_{y}H_{-\vk}^{*}\sigma_{y}=H_{\vk}$, where $\sigma_{y}$
acts on the spin, which yields the Hamiltonian structure above. When
$\mathcal{I}$-symmetry is also present, $H_{\vk}$ will be doubly degenerate, since the combined
operation $\mathcal {TI}$ leads to a Kramers pair at every momentum. $\mathcal{H}_{\vk}^{\textrm{BdG}}$ has particle hole symmetry implemented by
the transformation $\mathcal{C}=\Pi_{y}\sigma_{y}{\mathcal{K}}$,
where $\Pi$ matrices act on Nambu particle-hole indices, and ${\mathcal{K}}$
is complex conjugation. %A vortex along the $z$ direction, given by $\Delta(\vr)=|\Delta(r)|e^{-i\theta}$,
%where $r e^{i\theta} = x+iy$,
A vortex given by $\Delta(\vr)=|\Delta(r)|e^{-i\theta}$, breaks $\mathcal{T}$
but preserves $\mathcal{C}$. %\textbf{The combination of $\mathcal{T}$ and inversion symmetry  ensures that }

\begin{figure}[tb]
\begin{raggedright} \includegraphics[width=0.9\columnwidth]{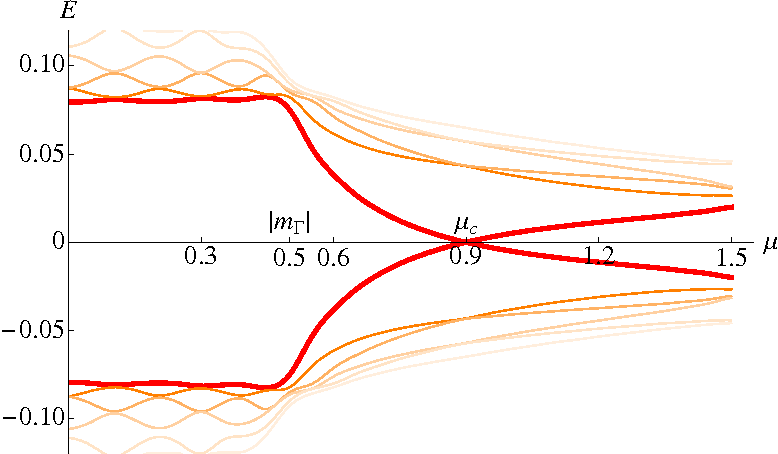}\\

\end{raggedright}

\begin{centering}
\includegraphics[bb=0bp 50bp 794bp 545bp,clip,width=1\columnwidth]{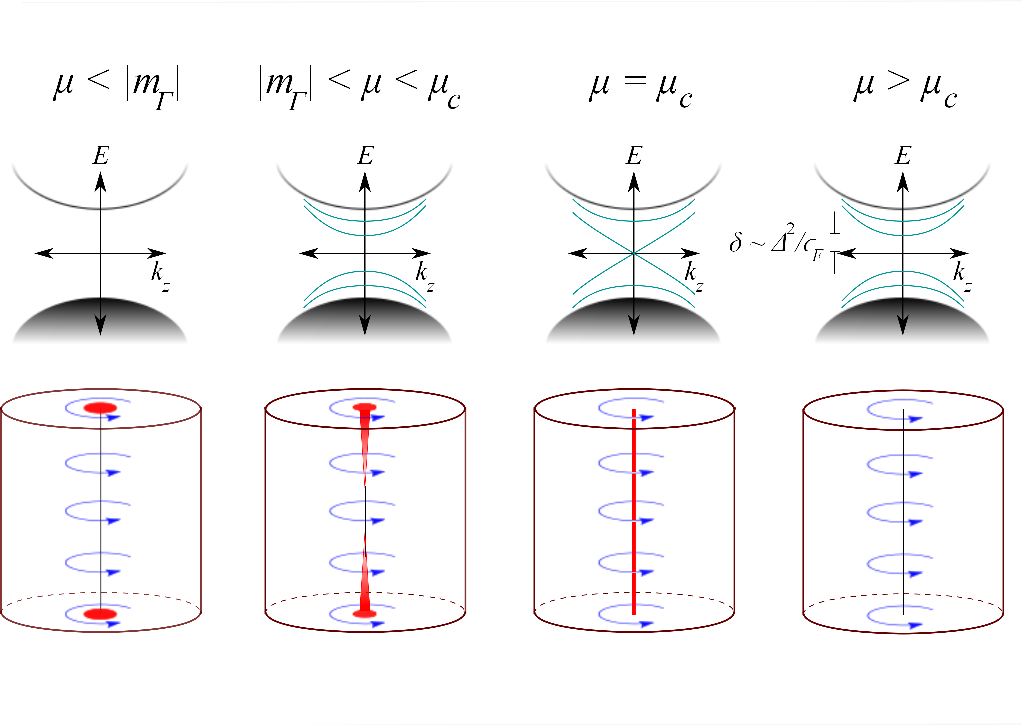}
\par\end{centering}

\caption{\emph{The vortex phase transition}. Evolution of the lowest bulk/vortex
states at $k_{z}=0$ (top row), of the dispersion within the vortex
(middle row) and of the surface MZMs (bottom row) as $\mu$ is varied
when the normal phase has a band inversion at $\Gamma$. At $\mu=0$,
the normal phase is a strong TI and a superconducting vortex traps
a MZM at its ends. As $\mu$ is increased, it first enters the conduction
band at $\mu=|m_{\Gamma}|$ and mid-gap states appear inside the vortex.
For $\mu<\mu_{c}$, the vortex stays gapped, but with a minigap $\delta$
smaller than the bulk gap. The MZMs remain trapped near the surface.
At $\mu_{c}=0.9$, the gap vanishes signalling a phase transition.
Beyond $\mu_{c}$, the vortex is gapped again, but there are no surface
MZMs. We used the lattice Hamiltonian with the parameters $t=0.5$,
$M=2.5$ and $m_{0}=-1.0$. The pairing strength is $\Delta_{0}=0.1$
far away from the vortex and drops sharply to zero at the core. Other
gap profiles give similar results.\label{fig:crossings}}

\end{figure}

{\em Role of vortex orientation:} %Inversion symmetry ($\mathcal{I}$) of the band structure of $H$ introduces an important simplification.
%A unit vortex passing through the inversion center will
%have $\Delta_{\vr}\rightarrow-\Delta_{\vr}$ under inversion. Invariance
%is readily restored by an additional gauge transformation of fermions
%by phase $\pi/2$, which defines the new inversion transformation.
Consider a straight vortex along $\hat z$. The dispersion $\mathcal{E}(k_{z})$
in general, has a minigap $\delta$ as in Fig.~\ref{fig:crossings}. A topological phase transition requires closing of the minigap and reopening with inverted sign. Since the $\mathbb{Z}_{2}$ topological
index is only changed by an odd number of such band crossings, the
only relevant momenta are $k_{z}=0,\,\pi$.
Band touchings at other $k_{z}$ points occur in pairs at $\pm k_{z}$
which do not alter the $\mathbb{Z}_{2}$ index ~\cite{KitaevWireMajorana01}.
In the weak pairing limit, one expects the critical point $\mu_{c}$
to be determined by a FS property, which will be outlined in detail
below. Here we simply observe that the relevant FSs to consider lie
in the $k_{z}=0,\,\pi$ planes, the planes determined by the vortex
orientation. %A different vortex orientation requires studying a different cross section of the FS.
This implies that the topological phase of the vortex, and hence $\mu_{c}$
depend in general on its orientation.

%In the weak pairing limit, the topological phase of the vortex must
%be related to a property of the FS before pairing was introduced.
%Below, we specialize to normal state band structures with both time
%reversal (${\mathcal{T}}$) and inversion (${\mathcal{I}}$) symmetry.
%For a vortex along the $z$ direction, we focus on FSs at $k_{z}=0$
%and $k_{z}=\pi$. Only level crossings at these momenta can change
%the $\mathbb{Z}_{2}$ topological index of the vortex.

\textbf{VPT in a lattice model:} Before discussing the general criterion
for a VPT, we present numerical and analytical evidence in a specific
lattice model from Ref.~\cite{HosurRyuChiralTISC}. While the numerics
explicitly demonstrate the phase transition, the analytical treatment
of the continuum limit allows us to conjecture a Berry phase condition
for the transition, which is later proved. The model is on a simple
cubic lattice with two orbitals per site: $H^\textrm{latt}_{\vk}=\tau_{x}\boldsymbol{d_{k}\cdot\sigma}+m_{\boldsymbol{k}}\tau_{z}-\mu$
where $\tau_{i}$ ($\sigma_{i}$) are Pauli matrices in the orbital
(spin) basis, $d_{\vk}^{i}=2t\sin k_{i}$, $m_{\vk}=\left(M+m_{0}\sum_{i}\cos k_{i}\right)$,
$i=x,y,z$, and $t$, $m_{0}$ and $M$ are parameters of the model
and $\mu$ is the chemical potential. %A strong topological insulator occurs if
The model is in the strong TI phase if $-3<\frac{M}{m_{0}}<-1$. %By varying $M$ and $m_{x,y,z}$,
%one can access all the topologically distinct insulators that preserve
%$\mathcal{T}$ and $\mathcal{I}$ in 3D.
%In the presence of a superconducting vortex along $z$, $k_{z}$ is conserved.
We add a mean field $s$-wave pairing to this Hamiltonian, insert
a unit winding into the pairing function and diagonalize the Hamiltonian
numerically. We focus on $k_{z}=0$.

{\em Numerical results:} Fig.~\ref{fig:crossings} illustrates
the evolution of the bulk vortex bound states, the dispersion within
the vortex and the surface MZMs as a function of $\mu$, when the
normal state has a band inversion only at the $\Gamma=(0,\,0,\,0)$-point,
\textit{i.e.}, $m_{\Gamma}<0$. At $\mu=0$, the bulk is gapped and
must have a pair of MZMs on opposite surfaces in a slab geometry.
As $\mu$ is raised, these MZMs leak deeper into the bulk, but survive
even after $\mu$ crosses $|m_{\Gamma}|$ despite the bulk now having
a FS in the normal phase, gapped by superconductivity. A VPT eventually
occurs at $\mu_{c}=0.9$, at which the vortex is gapless and the surface
MZMs merge into vortex line. Beyond $\mu_{c}$, there are no longer
any protected MZMs on the surface. %

{\em Continuum limit:} In the continuum limit of the lattice model, we can analytically calculate $\mu_{c}$. For
$k_{z}=0$ and small $k_{x,y}$ around $\Gamma$, $H^\textrm{latt}_{\vk}$
reduces to the isotropic form $\mathcal{H}_{\boldsymbol{k}}=v_{D}\tau_{x}\boldsymbol{\sigma\cdot k}+(m-\epsilon k^{2})\tau_{z}-\mu$. In this form, a band inversion exists if $m\epsilon>0$. Thus, $m\epsilon<0$
($>0$) defines a trivial insulator (strong TI). At $k=\sqrt{m/\epsilon}$,
$m_{\boldsymbol{k}}=m-\epsilon k^{2}$ vanishes and $\mathcal{H}_{\boldsymbol{k}}$
resembles two copies of a TI surface. In particular, the
Berry phase around each $\tau_{x}=\pm1$ FS is $\pi$. We show
later that this leads to a pair of vortex zero modes, signaling the
VPT at $\mu_{c}=v_{D}\sqrt{m/\epsilon}$.

We solve analytically for the two bulk zero modes at $\mu$ to first
order in $\Delta_{0}$ assuming $\left|\Delta(\vr)\right|=\Delta_{0}\Theta(r-R)$,
where $\Theta$ is the step-function and $\mu R/v_D\gg1$. Calculating the zero modes separately for
$r\leq R$ and $r\geq R$ and matching the solutions at $r=R$ gives a pair of zero
modes, only when $\mu=v_{D}\sqrt{m/\epsilon}$, for all vortex orientations. This is
precisely where the momentum dependent `mass' term changes sign~
\cite{SupplementaryMaterial}. Using the model parameters and the linearized approximation gives
an estimate of $\mu_{c}\approx1$, in agreement with the lattice
numerics. %close to what is seen numerically in the lattice mode.

\textbf{General Fermi surface Berry phase condition:} For weak pairing,
the VPT is expected to be governed by properties of the bulk FS. For
concreteness, we begin by assuming we have a {\em single} FS in the
$k_{z}=0$ plane, which will be doubly degenerate due to the combined symmetry
${\mathcal{TI}}$. We now argue that the VPT occurs when an appropriately defined Berry
phase for each of the two degenerate bulk FSs is $\pi$.

A convenient model for the vortex is $\Delta(\vr)=\frac{\Delta_{0}}{\xi}(x-iy)$.
The linear profile here simplifies calculations, but does not affect
location of the zero mode. %The choice of $\xi$ as the superconductor coherence length gives the right minigap scale for the low energy excitations.
The choice of $\xi$ as the length scale gives the right minigap scale
for the low energy excitations. Working in momentum space, we substitute
${\vr}$ by $i\partial_{\vk}$, which gives \begin{align}
\mathcal{H}_{\vk}^{\textrm{BdG}}=\begin{bmatrix}H_{\vk}-\mu & i\frac{\Delta_{0}}{\xi}(\partial_{k_{x}}-i\partial_{k_{y}})\\
i\frac{\Delta_{0}}{\xi}(\partial_{k_{x}}+i\partial_{k_{y}}) & \mu-H_{\vk}\end{bmatrix},\label{eq:HBdGk}\end{align}
 transforming now to the band basis $\ket{\varphi_{\vk}^{\nu}}$,
which are eigenstates of the band Hamiltonian $H_{\vk}\ket{\varphi_{\vk}}=E\ket{\varphi_{\vk}}$.
Since we are only interested in very low energy phenomena, we project
onto the two degenerate bands near the Fermi energy $\nu=1,2$. The
projected Hamiltonian then is: \begin{align}
\tilde{\mathcal{H}}_{\vk}^{\textrm{BdG}}=\begin{bmatrix}E_{\vk}-\mu & i\frac{\Delta_{0}}{\xi}(D_{k_{x}}-iD_{k_{y}})\\
i\frac{\Delta_{0}}{\xi}(D_{k_{x}}+iD_{k_{y}}) & -E_{\vk}+\mu\end{bmatrix}.\label{eq:HamIndicies}\end{align}
 where $D_{k_{\alpha}}=\partial_{k_{\alpha}}-i{\bf A}_{\alpha}(\vk)$
and ${\bf A}_{\alpha}(\vk)$, the $SU(2)$ connections, are 2$\times$2
matrices: $[{\bf A}]_{\alpha}^{\mu\nu}(\vk)=i\bra{\varphi_{\vk}^{\mu}}\partial_{k_{\alpha}}\ket{\varphi_{\vk}^{\nu}}$.
%The momenta $\vk$ are restricted to be near the FS in the $k_z=0$ plane.

{\em (i) Abelian case:} Let us first consider the case when an
additional quantum number (such as spin) can be used to
label the degenerate FSs. Then, $[{\bf A}]_{\alpha}^{\mu\nu}$ must
be diagonal, and reduces to a pair of $U(1)$ connections for the
two FSs. In this situation, (\ref{eq:HamIndicies}) is identical to
the effective Hamiltonian for a $p_{x}+ip_{y}$ SC, if
we interpret momenta as position and ignore the gauge potential. The
diagonal terms represent a transition from weak to strong pairing
phase on crossing the FS when $E_{\vk}=\mu$~\cite{ReadGreenP+ipFQHE00}.
Thus mid-gap are expected, composed of states near the Fermi energy.
Due to the finite size of the FS, these states have an energy spacing
of $\mathcal{O}\big(\frac{\Delta_{0}}{k_{F}\xi}\big)$, the minigap
energy scale. However, a zero energy state appears if the FS encloses
a $\pi$-flux \cite{ReadGreenP+ipFQHE00}. %\cite{ReadGreenP+ipFQHE00,BocquetZirnbauerDiracRandomMass00,ReadLudwigVortexDisorder00}.
This can be implemented via the gauge potential if $\oint_{_{\textrm{FS}}}\!{\mathbf{A}}\cdot\mathrm{d}{\mathbf{l}}=\pi$
leading to a pair of zero modes, since the other FS has the same Berry
phase by $\mathcal T$. %Note, the quantum numbers labeling the
%FSs should exchange under time reversal, so a nonzero phase can arise.

{\em (ii) General case, $SU(2)$ connection:} In the absence of
any quantum number distinguishing the bands, one integrates the vector
potential $\mathbf{A}(\vk)$ around the FS in the $k_{z}=0$ plane,
to give the non-Abelian Berry phase: $U_{B}=\mathcal{P}\exp\big[i\oint_{_{\textrm{FS}}}\!\mathbf{A}\cdot\mathrm{d}\mathbf{l}\big]\in SU(2)$,
where $\mathcal{P}$ denotes path ordering. (There is no $U(1)$ phase
by $\mathcal{T}$ symmetry.) Although $U_{B}$ itself depends on the
choice of basis, its eigenvalues $e^{\pm i\phi_{B}}$ are gauge invariant.
A semiclassical analysis~ \cite{SupplementaryMaterial} gives the Bohr-Sommerfield
type quantization condition for the low energy levels: \begin{equation}
E_{n}=\frac{\Delta_{0}}{l_{F}\xi}(2\pi n+\pi\pm\phi_{B})\end{equation}
 where $n$ is an integer and $l_{F}$ is the FS perimeter. A pair
of zero modes appears when $\phi_{B}=\pi$, \textit{i.e.} when $U_{B}=-\textbf{1}$.

We have considered a single closed FS in the $k_{z}=0$ plane. Such
a FS necessarily encloses a $\mathcal T$ invariant momentum (TRIM),
(\textit{e.g.} $\Gamma$), given the symmetries. When there are multiple
FSs, the condition above is applied individually to each FS, since
tunneling between them is neglected in the semiclassical approximation.
Closed FSs that do not enclose a TRIM, or pairs of open FSs, cannot
change the vortex topology.

\textbf{Candidate materials:} We now apply the Berry phase criterion
to some candidate materials to see which of them can have protected
MZMs at the ends of vortices.

{\em Cu$_{x}$Bi$_{2}$Se$_{3}$:} The insulating phase of Bi$_{2}$Se$_{3}$
is a strong TI with a single band inversion occurring at the $\Gamma$
point. On Cu doping, Bi$_{2}$Se$_{3}$ becomes $n$-type with an
electron pocket at $\Gamma$ and is reported to superconduct below
$T_{c}=\unit[3.8]{K}$~\cite{SuperconductingCuxBi2Se3,SCdoping}.
%Cu creates an electron pocket at $\Gamma$ in Bi$_{2}$Se$_{3}$
%and the carrier density just before the superconducting
%transition is approximately $2\times10^{20}\mbox{cm}^{-3}$~\cite{SCdoping} from Hall measurement{\bf Photoemission measurements show $\mu\approx\unit[0.25]{eV}$ above the conduction band minimum at optimal doping~\cite{CuxBi2Se3ARPES}.
Photoemission measurements show $\mu\approx\unit[0.25]{eV}$ above
the conduction band minimum at optimal doping ~\cite{CuxBi2Se3ARPES}.
%we estimate $\mu$ theoretically in this material from the carrier
%density to be $\approx0.4eV$ relative to the conduction band bottom.
%Using the same model, we determine $\mu_{c}$ for this material by
We calculate the Berry phase eigenvalues for a FS around the $\Gamma$
point numerically as a function of $\mu$, which evaluates to $\pm\pi$
at $\mu_{c}\approx\unit[0.24]{eV}$ above the conduction band minimum
for a vortex along the $c$-axis of the crystal~\cite{SupplementaryMaterial}. Hence $\mu\gtrsim\mu_{c}$                 
indicates c-axis vortices are near the topological transition\cite{SupplementaryMaterial}. However, tilting the vortex away from the c-axis is found to raise $\mu_c$ to upto $\mu_c=0.30eV$, for a vortex perpendicular to the c-axis. Therefore, sufficiently tilted vortices should host MZMs at the experimental doping level.

{\em p-doped TlBiTe$_{2}$, p-doped Bi$_{2}$Te$_{3}$ under pressure
and Pd$_{x}$Bi$_{2}$Te$_{3}$: }The bands of TlBiTe$_{2}$ and
Bi$_{2}$Te$_{3}$ are topologically non-trivial because of a band
inversion at the $\Gamma$ point \cite{TlBiX2predictionSCZhang,*TlBiX2predictionHasan,*TlBiX2predictionChulkov,*TlBiX2experimentShen,*ChenBi2Te3,*SCZhangBi2X3DiracCone}.
The topological character of Bi$_{2}$Te$_{3}$ is believed to be
preserved under a pressure of up to $\unit[6.3]{GPa}$, at which it
undergoes a structural phase transition. On $p$-doping to a density
of $\unit[6\times10^{20}]{cm^{-3}}$ ($\unit[3\mbox{-}6\times10^{18}]{cm^{-3}}$),
TlBiTe$_{2}$ (Bi$_{2}$Te$_{3}$ under $\unit[3.1]{GPa}$) becomes
a SC below $T_{c}=\unit[0.14]{K}$ $(\sim\!\unit[3]{K})$~\cite{TlBiTe2Superconductivity,ZhangBi2Te3SCunderPressure},
making it a natural system to search for the possibility of MZMs.
Similarly, $n$-doping Bi$_{2}$Te$_{3}$ to a concentration of $\unit[9\times10^{18}]{cm^{-3}}$
by adding Pd reportedly results in $T_{c}=\unit[5.5]{K}$~\cite{SCdoping}
in a small sample fraction. The superconductivity in Bi$_{2}$Te$_{3}$
under pressure, and in TlBiTe$_{2}$ (Pd$_{x}$Bi$_{2}$Te$_{3}$)
is believed to arise from six symmetry related hole (electron) pockets
around the $\Gamma$-$T$ line. % which are symmetry related.
%to each other by the threefold rotational symmetry of the lattice
%about the $c$-axis and by spatial inversion. Because of inversion
%symmetry, the total number of VPTs as $\mu$ is hypothetically varied
%from inside the band gap to the experimental value can only be even.
This is an even number so vortex lines in superconducting TlBiTe$_{2}$
and both $p$- and $n$-type Bi$_{2}$Te$_{3}$ should have MZMs at
their ends in all orientations.

{\em MZMs from trivial insulators:} The bulk criterion derived
does not require a `parent' topological band structure. As a thought
example, say we have four TRIMs with Hamiltonians like the continuum Hamiltonian $\mathcal{H}_{\boldsymbol{k}}$ in their vicinity. Such band inversions at four TRIMs in a plane leads
to a trivial insulator \cite{FuKaneTIInversion}. However, if their
critical chemical potentials $\mu_{c}$ differ, then there could be a range of $\mu$ where there are an \emph{odd} number of
VPTs below and above $\mu$, leading to topologically non-trivial
vortices. %If say three are related by symmetry, and one is distinct, it is readily seen that there is a range of chemical potentials where vortices are in the topological phase.
Interestingly PbTe and SnTe are both trivial insulators with band
inversions relative to each other at the four equivalent $L$ points.
They both exhibit superconductivity on doping below $T_{c}=\unit[1.5]{K}$~\cite{PbTeSuperconductivity}
and $\unit[0.2]{K}$~\cite{HeinSnTeSC} respectively. A combination
of strain (to break the equivalence of the four $L$ points) and doping
could potentially create the scenario described above in one of these
systems. GeTe is similar to SnTe with $T_{c}\sim\unit[0.3]{K}$~\cite{GeTeSuperconductivity}
but undergoes a spontaneous rhombohedral distortion resulting in the
desired symmetry. Thus, $\mathcal I$- and $\mathcal T$-symmetric systems with strong spin orbit can lead to SCs with vortex end MZMs, even in the absence of a proximate topological phase. Investigating the Fermi surface SU(2) Berry phases, and thus the vortex electronic structure,  in this wide class of systems is a promising future direction in the hunt for Majorana fermions. 

In closing, we note that the VPT could potentially be probed via thermal
transport along the vortex line. A hurdle is the small minigap scale ($\Delta/k_{F}\xi\sim\Delta^{2}/E_{F}$),
and the long confinement length of the MZMs to the surface,
which may be ameliorated by considering strong coupling SCs
or materials such as heavy fermions where $E_{F}$ is 
reduced.

%{\em PbTe, SnTe and GeTe:} PbTe (SnTe) are trivial insulators on
%a FCC (rhombohedral) lattice with very similar band structures. On
%$p$-doping, PbTe (SnTe) shows superconductivity below .
%An important difference between the band structures is that PbTe has
%band inversions at four TRIMs relative to SnTe\cite{TlBiX2predictionHasan}.
%The four TRIMs are the four $L$-points in PbTe and the three $L$
%and the $T$ points in SnTe. Close to these four points, one of these
%materials has a mass gap that changes sign as one moves away from
%the TRIM. If the mass inversion occurs in SnTe, the $L$ and $T$
%points will have different $\mu_{c}$'s in general for a VPT in a
%certain direction. It may be possible to tune $\mu$ to lie between
%the two $\mu_{c}$'s while preserving the superconductivity. A vortex
%in this SC would then host protected MZMs at its ends. If, on the
%other hand, PbTe has the band inversions, a similar result may be
%achieved by breaking the symmetry relating the four $L$ points by
%the vortex direction or by applying a strain. GeTe has a band structure
%similar to that of SnTe and is also known to superconduct below $\sim0.1K$\cite{SnTeGeTeBandStructure,GeTeSuperconductivity}.
%This raises the exciting prospect of obtaining a topologically non-trivial
%superconducting vortex starting from a material with a trivial band
%structure which merits further investigation.

We thank A.~M. Turner, J.~H. Bardarson,
A. Wray and C.~L. Kane for insightful discussions, and NSF-DMR 0645691 for funding. In parallel
work, L. Fu, J.~C.~Y. Teo and C.~L. Kane have arrived at similar conclusions.

\bibliography{FullDraft}

\end{document}

\title{Supplementary material for `Majorana modes at the ends of superconductor vortices in doped topological
insulators'}

\author{Pavan Hosur}

\affiliation{Department of Physics, University of California, Berkeley, CA 94720,
USA}

\author{Pouyan Ghaemi}

\affiliation{Department of Physics, University of California, Berkeley, CA 94720,
USA}

\affiliation{Materials Sciences Division, Lawrence Berkeley National Laboratory,
Berkeley, CA 94720}

\author{Roger S.~K. Mong}

\affiliation{Department of Physics, University of California, Berkeley, CA 94720,
USA}

\author{Ashvin Vishwanath}

\affiliation{Department of Physics, University of California, Berkeley, CA 94720,
USA}

\affiliation{Materials Sciences Division, Lawrence Berkeley National Laboratory,
Berkeley, CA 94720}

\maketitle

\section{Continuum model: Analytical solution}

In the continuum limit of the lattice model with cubic symmetry, we
can analytically calculate $\mu_{c}$. For $k_{z}=0$ and small $k_{x,y}$
around $\Gamma$, the lattice model reduces to the isotropic form
$\mathcal{H}_{\boldsymbol{k}}=v_{D}\tau_{x}\boldsymbol{\sigma\cdot k}+(m-\epsilon k^{2})\tau_{z}-\mu$.
In this form, a band inversion exists if $m\epsilon>0$. Thus, $m\epsilon>0(<0)$
characterizes a strong TI (trivial insulator). At $k=\sqrt{m/\epsilon}$,
$m_{\boldsymbol{k}}=m-\epsilon k^{2}$ vanishes. Below we show that
$\mu_{c}=v_{D}\sqrt{m/\epsilon}$ is the critical chemical potential
at which the VPT occurs.

We solve for the two bulk zero modes at $\mu_{c}$ analytically to
first order in $\Delta_{0}$ by assuming $\left|\Delta(\boldsymbol{r})\right|=\Delta_{0}\Theta(r-R)$,
where $\Theta$ is the step-function $R$ is a large radius such that
$R(\Delta_{0}/v_{D}\hbar)\gg1$. In the basis $\left(\psi(\boldsymbol{r}),\, i\sigma_{y}\psi^{\dagger}(\boldsymbol{r})\right)$
where $\psi(\boldsymbol{r})$ is a four-component spinor of electron
annihilation operators indexed by spin and orbital indices, the Bogoliubov-de
Gennes Hamiltonian in real space for a unit vortex is \begin{equation}
\mathcal{H}^{\textrm{BdG}}=\left(\begin{array}{cc}
\mathcal{H}_{\boldsymbol{r}} & \Delta_{0}\Theta(r-R)e^{i\theta}\\
\Delta_{0}\Theta(r-R)e^{-i\theta} & -\mathcal{H}_{\boldsymbol{r}}\end{array}\right)\end{equation}
 where $\mathcal{H}_{\boldsymbol{r}}=-iv_{D}\tau_{x}\boldsymbol{\sigma\cdot\nabla}+\tau_{z}(m+\epsilon\nabla^{2})-\mu$.
The $\theta$-dependence of $\mathcal{H}^{\textrm{BdG}}$ can be removed
by observing that it commutes with the generalized angular momentum
operator $L_{z}=-i\partial_{\theta}+\frac{\sigma_{z}+\Pi_{z}}{2}$.
Since $\left\{ L_{z},\mathcal{C}\right\} =0$, a single MZM, such
as the one on the surface, must have an eigenvalue $n$ of $L_{z}$,
equal to zero. Now, the bulk zero modes can be thought of as the avenues
through which the surface MZMs penetrate the bulk. Thus, they must
have $n=0$ as well. The vortex preserves a mirror symmetry about
the $xy$-plane described by $\mathcal{M}=\tau_{z}\sigma_{z}$. Thus,
$\mathcal{H}^{\textrm{BdG}}$ can be block-diagonalized into sectors
with opposite $\mathcal{M}$ eigenvalues. The two blocks are particle-hole
conjugates since $\left\{ \mathcal{M},\mathcal{C}\right\} =0$ and
each must contribute a single bulk MZM at $\mu_{c}$. The radial Hamiltonian
for the $\mathcal{M}=+1$ sector is \begin{widetext} \begin{eqnarray}
\mathcal{H}_{\textrm{rad}} & = & -i\Pi_{z}\nu_{y}\left(\partial_{r}+\frac{1}{2r}\right)+\frac{\nu_{x}}{2r}+\Pi_{z}\nu_{z}\left[m+\epsilon\left(\partial_{r}^{2}+\frac{1}{r}\partial_{r}-\frac{1}{2r^{2}}\right)\right]-\frac{\epsilon}{2r^{2}}-\Pi_{z}\mu+\Pi_{x}\Delta_{0}\Theta(r-R)\end{eqnarray}
 \end{widetext} where $\nu_{i}$ are Pauli matrices in a combined
orbital $(\tau_{i})$ and spin $(\sigma_{i})$ space. For $r<R$,
$\mathcal{H}_{\textrm{rad}}$ has four zero modes \begin{align}
\begin{pmatrix}-k_{\pm}J_{1}(k_{\pm}r)\\
(\mu-m_{\boldsymbol{k}_{\pm}})J_{0}(k_{\pm}r)\\
0\\
0\end{pmatrix}, & \begin{pmatrix}0\\
0\\
-k_{\pm}J_{0}(k_{\pm}r)\\
(\mu-m_{\boldsymbol{k}_{\pm}})J_{1}(k_{\pm}r)\end{pmatrix},\end{align}
 where $k_{\pm}$ are roots of $g(k)=\epsilon^{2}k^{4}+(1-2m\epsilon)k^{2}+(m^{2}-\mu^{2})$
and $J_{i}(x)$ are Bessel functions of the first kind of order $i$.
For $r>R$, we drop all terms that contain $r$ in the denominator.
The zero modes of the remaining Hamiltonian are of the form $e^{i\lambda_{\pm}r}\chi_{\pm}$,
where $\lambda_{\pm}$ is a root of $\epsilon^{2}\lambda^{4}+(1-2m\epsilon)\lambda^{2}+\left(m^{2}-(\mu\pm i\Delta_{0})^{2}\right)=0$
with positive imaginary part and \begin{align}
\chi_{\pm}=\begin{pmatrix}-i\lambda_{\pm}\\
\mu\mp i\Delta_{0}-m+\epsilon\lambda_{\pm}^{2}\\
\pm\lambda_{\pm}\\
\pm i(\mu\mp i\Delta_{0}-m+\epsilon\lambda_{\pm}^{2})\end{pmatrix}.\end{align}
 Matching the solutions and their derivatives at $r=R$ to first order
in $\Delta_{0}$, using the asymptotic forms of the Bessel functions,
is only possible at $\mu=v_{D}\sqrt{m/\epsilon}$. This gives an analytic
solution to the critical chemcial potential $\mu_{c}$.

\section{Vortex modes and the Berry phase of the Fermi surface}

We consider a 3D system with $s$-wave pairing, where the bulk is
pierced by a quantum flux $h/2e$. Assuming that the metallic Hamiltonian
$H_{\vk}$ (the system without superconductivity/quantum flux) has
time-reversal symmetry, we will derive the condition which governs
the existence of zero vortex modes to the properties of the FS, namely
the Berry phase.

The mean field Bogoliubov-de Gennes (BdG) Hamiltonian with $s$-wave
pairing is of the form: \begin{align}
\mathcal{H} & =\frac{1}{2}\sum_{\vk\vk'}\Psi_{\vk}^{\dag}\mathcal{H}^{\textrm{BdG}}(\vk,\vk')\Psi_{\vk'}^{\phd},\end{align}
 where $\Psi_{\vk}^{\dag}=\big(\mathbf{c}_{\vk}^{\dag},\mathbf{c}_{-\vk}^{T}(i\sigma_{y})\big)$
is written in the Nambu basis, capturing all orbital and spin degrees
of freedom. The single particle Hamiltonian is \begin{align}
\mathcal{H}^{\textrm{BdG}}(\vk,\vk')=\begin{bmatrix}(H_{\vk}-\mu)\delta_{\vk\vk'} & \Delta(\vr)\\
\Delta^{\ast}(\vr) & (\mu-\sigma_{y}H_{-\vk}^{\ast}\sigma_{y})\delta_{\vk\vk'}\end{bmatrix},\end{align}
 where the pairing potential, given by $\Delta^{\ast}(\vr)=\big\langle\Psi_{\uparrow}^{\dag}(\vr)\Psi_{\downarrow}^{\dag}(\vr)\big\rangle$,
is position dependent due to the vortex.

The quasi-particle operators $\Psi^{\dag},\Psi$ are defined in such
as a way that leads to spin singlet pairing. Since the metallic Hamiltonian
$H_{\vk}$ has time-reversal symmetry, $\sigma_{y}H_{-\vk}^{\ast}\sigma_{y}=H_{\vk}$
and the BdG Hamiltonian may be written as \begin{align}
\mathcal{H}^{\textrm{BdG}}=\begin{bmatrix}H_{\vk}-\mu & \Delta(\vr)\\
\Delta^{\ast}(\vr) & \mu-H_{\vk}\end{bmatrix}.\label{eq:sWaveVortexHam}\end{align}
 (We drop the $\vk,\vk'$ dependence, treating $\mathcal{H}^{\textrm{BdG}}$
as an operator.)

We take the vortex to lie in the $\hat{z}$ direction, hence the pairing
term takes the form $\Delta^{\ast}(\vr)=\Delta(r_{\perp})e^{i\theta}$,
independent of $z$, where $r_{\perp}e^{i\theta}=x+iy$. The pairing
amplitude becomes constant for large $r_{\perp}$: $\Delta(r_{\perp})\rightarrow\Delta_{0}$.
Although the vortex breaks translational symmetry in the $xy$-plane,
it is preserved in the $z$ direction and hence, $k_{z}$ remains
a good quantum number. The 3D system thus decouples into many 2D Hamiltonians
enumerated by $k_{z}$. Henceforth, we refer to $r_{\perp}$ as simply
$r$, and $\vk$ as the 2D momentum coordinate $(k_{x},k_{y})$.

The time-reversal operator is $-i\sigma_{y}\mathcal{K}$, taking $k_{z}\rightarrow-k_{z}$,
$\vk\rightarrow-\vk$, which is broken by the imaginary part of $\Delta$.
In addition, the system must have particle-hole (charge conjugation)
symmetry given by the operator $\mathcal{C}=\Pi_{y}\sigma_{y}\mathcal{K}$,
where $\Pi_{i}$ are the Pauli matrices acting on particle-hole space.
Note that $\mathcal{C}$ also takes $k_{z}\rightarrow-k_{z}$, $\vk\rightarrow-\vk$,
hence particle-hole is a symmetry of the 2D system only when $k_{z}\sim-k_{z}$
(\textit{i.e.} at $0$ or $\pi$).

In the remainder of this section, we do not assume anything about
the value of $k_{z}$, nor the symmetries of the 2D Hamiltonian $H_{\vk}|_{k_{z}}$
at a fixed $k_{z}$. Our results remain valid applied to any 2D slice
(with a smooth FS) of the 3D Brillouin zone, as long as the 3D Hamiltonian
has time-reversal symmetry.

\subsection{Pairing potential}

In this section, we compute the pairing potential of a vortex in $k$-space
via a Fourier transform. %\begin{align}
%	\Delta^\ast(\vr) & = \xp{ \Psi_{\uparrow}^\dag(\vr) \Psi_{\downarrow}^\dag(\vr) }
%		= \Delta(r)e^{i\theta} .
%\end{align}
We confine our system to be on a disk with radius $\xi$, the superconducting
correlation length. The matrix element $\Delta_{\vk'\vk}$ is \begin{align*}
\Delta_{\vk'\vk} & =\bra{\vk'}\Delta(\vr)\ket{\vk}\\
 & =\frac{1}{\pi\xi^{2}}\int\! d^{2}r\, e^{i(\vk-\vk')\cdot\vr}\Delta(\vr)\\
 & =\frac{1}{\pi\xi^{2}}\int_{0}^{\xi}\!\! rdr\,\Delta(r)\int_{0}^{2\pi}\!\! d\theta\, e^{iqr\cos(\theta-\theta_{q})}e^{-i\theta},\end{align*}
 where $\vq=\vk-\vk'=q(\cos\theta_{q}\hat{x}+\sin\theta_{q}\hat{y})$.
The $\theta$ integral evaluates to a Bessel function (of the first
kind): $2\pi ie^{-i\theta_{q}}J_{1}(qr)$. The matrix element becomes
\begin{align}
\Delta_{\vk'\vk} & =\frac{2\pi ie^{-i\theta_{q}}}{\pi\xi^{2}}\int_{0}^{\xi}\!\! rdr\,\Delta(r)J_{1}(qr).\end{align}

\begin{itemize}
\item At large $q$ ($q\xi\gg1$), $J_{1}(qr)\approx\sin{qr}/\sqrt{qr}$,
and $\Delta(r)\rightarrow\Delta_{0}$ becomes constant. The integral
scales as $\frac{\Delta_{0}}{q^{2}}(q\xi)^{1/2}\cos q\xi$ and the
matrix element $\Delta_{\vk'\vk}\approx i\Delta_{0}e^{-i\theta_{q}}(q\xi)^{-3/2}\cos q\xi\rightarrow0$. 
\item At small $q$ ($q\xi<1$), $J_{1}(qr)\approx qr/2$ and the integral
evaluates to $q\Delta_{0}\xi^{3}/6$. The matrix element scales as
$\frac{i}{3}\Delta_{0}e^{-i\theta_{q}}q\xi$ and is linear in $q$. 
\end{itemize}
\begin{figure}[th]
 \includegraphics[width=0.45\textwidth,keepaspectratio=true]{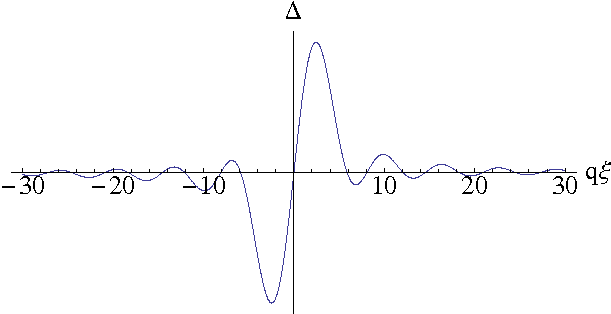} \caption{The pairing potential. $\Delta_{\vk'\vk}$ as a function of $q\xi$
where $q=|\vk-\vk'|$ and $\xi$ is the system radius. The wavefunction
$\psi(\vk)$ is smooth on the scale of $\xi^{-1}$, and hence the
pairing potential may be modeled as $-\delta'(q)$. }

\end{figure}

%When the characteristic momentum transfer is of the order
We can model the pairing matrix element as a derivative of the delta
function: $\Delta_{\vk'\vk}\propto ie^{-i\theta_{q}}(-\delta'(q))$,
this approximation is valid in the regime where the wavefunction $\psi(\vk)$
is smooth on the length scale of $\xi^{-1}$. In Cartesian coordinates,
the pairing term becomes $i\Delta_{e}(\partial_{k_{x}}-i\partial_{k_{y}})$,
where $\Delta_{e}$ is the effective $p+ip$ pairing strength, with
units energy $\times$ length$^{-1}$. From a simple analysis, we
expect that $\Delta_{e}\approx\Delta_{0}/\xi$.

The Hamiltonian in $k$-space is of the form: \begin{align}
\mathcal{H}^{\textrm{BdG}}=\begin{bmatrix}H_{\vk}-\mu & i\Delta_{e}(\partial_{k_{x}}-i\partial_{k_{y}})\\
i\Delta_{e}(\partial_{k_{x}}+i\partial_{k_{y}}) & \mu-H_{\vk}\end{bmatrix}.\label{eq:HamBdGkdk}\end{align}
 %Here $\Delta_e$ is the effective $p+ip$ pairing strength, with units energy $\times$ length$^{-1}$.
%A simple analysis shows that $\Delta_e \xi = \Delta_0 \ln(\xi/r_c)$.
The eigenstates are of the form \begin{align}
\ket{\psi}=\sum_{\vk,\nu}\psi_{\nu}(\vk)\ket{\varphi_{\vk}^{\nu}},\end{align}
 where $\ket{\varphi_{\vk}^{\nu}}$ is an eigenstate of $H_{\vk}$
and $\nu$ labels the band index.

\subsection{Projecting to the low energy states}

Solutions to the Hamiltonian \eqref{eq:HamBdGkdk} will consists of
mostly states near the FS, where $E_{\vk}\sim\mu$. Hence we can simplify
the system by projecting to the band eigenstates $\ket{\varphi_{\vk}^{\mu}}$
with energy near the chemical potential, where $\mu,\nu$ is the band
index: \begin{align}
\tilde{\mathcal{H}}=\bra{\varphi_{\vk}^{\mu}}\mathcal{H}\ket{\varphi_{\vk}^{\nu}}.\end{align}
If inversion symmetry is present, each band will be doubly degenerate.

The projection of the metallic Hamiltonian $H_{\vk}$ gives a diagonal
matrix $E_{\vk}^{\mu\nu}=\bra{\varphi_{\vk}^{\mu}}H_{\vk}\ket{\varphi_{\vk}^{\nu}}$
with its entries being the energies. The projection of the derivative
operator gives the Berry connection: $\bra{\varphi_{\vk}^{\mu}}i\nablak\ket{\varphi_{\vk}^{\nu}}=i\nablak+\mathbf{A}^{\mu\nu}$,
where the non-Abelian Berry connection is defined as: $\mathbf{A}^{\mu\nu}=i\bra{\varphi_{\vk}^{\mu}}\bm{\nablak}\ket{\varphi_{\vk}^{\nu}}$.
Explicitly: \begin{align}
\tilde{\mathcal{H}}=\begin{bmatrix}E_{\vk}^{\mu\nu}-\mu & \Delta_{e}\begin{pmatrix}i(\partial_{p_{x}}-i\partial_{p_{y}})\\
+A_{x}^{\mu\nu}-iA_{y}^{\mu\nu}\end{pmatrix}\\
\Delta_{e}\begin{pmatrix}i(\partial_{p_{x}}+i\partial_{p_{y}})\\
A_{x}^{\mu\nu}+iA_{y}^{\mu\nu}\end{pmatrix} & -E_{\vk}^{\mu\nu}+\mu\end{bmatrix}.\label{eq:p+ipHamIndicies}\end{align}
 The connections $A_{x},A_{y}$ are hermitian matrices which depends
on the choice of Bloch states $\ket{\varphi_{\vk}^{\nu}}$.

%Explicitly, the Hamiltonian becomes
Dropping the band indices and writing $E_{\vk}$ as $E(\vk)$, the
Hamiltonian is: \begin{align}
\tilde{\mathcal{H}} & =\begin{bmatrix}E(\vk)-\mu & \Delta_{e}\begin{pmatrix}i(\partial_{p_{x}}-i\partial_{p_{y}})\\
+A_{x}-iA_{y}\end{pmatrix}\\
\Delta_{e}\begin{pmatrix}i(\partial_{p_{x}}+i\partial_{p_{y}})\\
+A_{x}+iA_{y}\end{pmatrix} & -E(\vk)+\mu\end{bmatrix},\label{eq:p+ipHam}\end{align}
 If we intepret momenta ($\vk$) as position, then we can treat $A_{x,y}$
as a gauge potential and the Berry curvature $F=\partial_{k_{x}}A_{y}-\partial_{k_{y}}A_{x}-i[A_{x},A_{y}]$
as an effective magnetic field in the model. This system was studied
by Read, Green, Ludwig, Bocquet and Zirnbauer in context of a Dirac
Hamiltonian with random mass as well as $p+ip$ superconducting systems~\cite{ReadGreenP+ipFQHE00,ReadLudwigVortexDisorder00,BocquetZirnbauerDiracRandomMass00}.

While superficially similar to a $p+ip$ superconductor, there is
an important distinction -- our system may not have particle-hole
symmetry (unless $k_{z}=0$ or $\pi$), and hence does not belong
to the D class. To illustrate how $\mathcal{C}$ is broken, suppose
that there is only one FS and hence $A_{i}$ are simply $1\times1$
(Abelian) matrices. The system resembles a superconductor with $p+ip$
pairing, but with an effective magnetic field $\bm{\nabla}\times\mathbf{A}$.
Unlike a superconductor, the effective field does not have to be localized
nor quantized within the FS.

Consequently, the spectrum of vortex states in our 2D system does
not have any symmetry, and the zero modes are not topologically protected.
We can only show that these modes are stable within the weak pairing
limit under perturbation theory. We can restore particle-hole symmetry
by combining the 2D systems at $k_{z}$ with that at $-k_{z}$, at
the cost of doubling the number of zero energy states.

%In particular, when a symmetry exist which separates the two FSs in to time-reversal conjugate pairs, our system decomposes in to two separate $p+ip$ Hamiltonians for each of the surfaces.  However, individually there is no particle-hole symmetry, as the symmetry takes one subsystem to the other.  As a result, each half of the system may have an arbitrary Berry phase ($e^{i\phi} = \mathcal{P}\exp\oint\! \mathbf{A}\cdot d\mathbf{l}$) that is not quantized, while the total system has zero net phase ($\operatorname{Tr}(\phi) = 0$).

In the remainder of the section, we will explicitly show the following:
when (an eigenvalue of) the Berry phase of the FS $\phi_{F}$ is $\pi$,
there is an effective half quantum flux $\frac{h}{2e}$ in the system
which supports a Majorana mode \cite{ReadGreenP+ipFQHE00,ReadLudwigVortexDisorder00,BocquetZirnbauerDiracRandomMass00}.
%The following sections will show this explicitly.

\subsection{Explicit solution with rotational symmetry}

This section is not necessary to the solution, but is instructive
and aids in the understanding of what the terms in the more general
solutions mean. For simplicity, we only consider a single FS at wavevector
$k_{F}$. We also assume an Abelian Berry connection, so $A_{x},A_{y}$
are simply real numbers.

With rotational symmetry, we can simplify the expressions in polar
coordinates: $k_{x}+ik_{y}=ke^{i\theta}$: $E(\vk)=E(k)$, $\partial_{k_{x}}-i\partial_{k_{y}}=e^{-i\theta}(\partial_{k}-\frac{i}{k}\partial_{\theta})$,
$A_{x}-iA_{y}=e^{-i\theta}(A_{k}-\frac{i}{k}A_{\theta})$. In addition,
it is possible to find a gauge for which $A_{k}=0$, and $A_{\theta}$
is a function of $k$, but independent of $\theta$. Explicitly, the
Berry connection is \begin{align}
2\pi A_{\theta}(k)=2\pi\int_{0}^{k}\! k'dk'\, F(k'),\label{eq:CircularAF}\end{align}
 where $F(\vk)=\nablak\times\mathbf{A}$ is the Berry curvature. The
left side of \eqref{eq:CircularAF} is the Berry phase along a circle
of radius $k$, while the right side is the integrated Berry curvature.

Our Hamiltonian simplifies to \begin{align}
\tilde{\mathcal{H}} & =\begin{bmatrix}E(k)-\mu & i\Delta_{e}e^{-i\theta}\left(\partial_{k}-\frac{i\partial_{\theta}+A_{\theta}}{k}\right)\\
i\Delta_{e}e^{i\theta}\left(\partial_{k}+\frac{i\partial_{\theta}+A_{\theta}}{k}\right) & -E(k)+\mu\end{bmatrix}.\end{align}
 The Hamiltonian commutes with $J_{z}=-i\partial_{\theta}+\Pi_{z}$,
that is to say that the solutions are of the form \begin{align}
\psi(k,\theta)=\frac{1}{\sqrt{k}}\begin{pmatrix}u(k)e^{i(n-1)\theta}\\
-iv(k)e^{in\theta}\end{pmatrix},\label{eq:CircularWf}\end{align}
 for integers $n$ (required by the wavefunction being single-valued).
Via the transformation \begin{align}
W_{ph}=k^{\frac{1}{2}}\begin{bmatrix}e^{-i(n-1)\theta}\\
 & i\, e^{-in\theta}\end{bmatrix},\label{eq:CircularWph}\end{align}
 the effective Hamiltonian for $(u,v)$ is \begin{align}
\tilde{\mathcal{H}}_{n} & =W_{ph}\,\tilde{\mathcal{H}}\, W_{ph}^{-1}\notag\\
 & =\begin{bmatrix}E(k)-\mu & \Delta_{e}\left(\partial_{k}+\frac{n-\frac{1}{2}-A_{\theta}}{k}\right)\\
\Delta_{e}\left(\frac{n-\frac{1}{2}-A_{\theta}}{k}-\partial_{k}\right) & -E(k)+\mu\end{bmatrix}.\label{eq:CircularWHamWdag}\end{align}
 Notice that the Hamiltonian is symmetric except for the terms proportional
to $\partial_{k}$, due to our choice of $W_{ph}$.

Our assumption is to replace $k$ by the Fermi wavevector $k_{F}$,
$A_{\theta}$ by $A_{\theta}^{F}=A_{\theta}(k_{F})$ and $\frac{1}{k}$
by $\frac{1}{k_{F}}$. This is justified as the amplitude of the wavefunction
$|u(k)|,|v(k)|$ is largest at the FS $k=k_{F}$ and exponentially
decays away from the FS. The resulting Hamiltonian is equivalent to
the Jackiw-Rebbi model~\cite{JackiwRebbi76}: \begin{multline}
\tilde{\mathcal{H}}_{n}=\frac{\Delta_{e}}{k_{F}}\left(n-A_{\theta}^{F}-\tfrac{1}{2}\right)\Pi_{x}+i\Delta_{e}\partial_{k}\Pi_{y}\\
+(E(k)-\mu)\Pi_{z}.\label{eq:PolarHamJackiwRabbi}\end{multline}
 A midgap state exists whenever $E(k)-\mu$ changes sign, with energy
\begin{align}
\mathcal{E}_{n} & =\frac{\Delta_{e}}{k_{F}}\left(n-\frac{\phi_{F}}{2\pi}-\frac{1}{2}\right),\label{eq:CircularVortexEnergies}\end{align}
 where $\phi_{F}=2\pi A_{\theta}^{F}$ is the Berry phase of the FS.
Hence, a zero energy solution exists when $\phi_{F}$ is an odd multiple
of $\pi$.

Explicitly, the eigenstates are of the form \cite{JackiwRebbi76,ReadGreenP+ipFQHE00}
\begin{align}
u(k)=\exp\int^{k}\!\beta(k')\, dk',\end{align}
 where $\beta(k)$ must a be a decreasing function of $k$ for $u(k)$
to be normalizable. Assuming that $E(k)$ is an increasing function
of $k$, then \begin{subequations} \begin{align}
\beta(k) & =\frac{\mu-E(k)}{\Delta_{e}}\\
v(k) & =u(k)\end{align}
 \end{subequations} By inspection, this satisfies the Schrödinger
equation for the Hamiltonian \eqref{eq:PolarHamJackiwRabbi}: \begin{align}
\tilde{\mathcal{H}}_{n}\begin{pmatrix}u\\
v\end{pmatrix}=\begin{bmatrix}-\Delta_{e}\beta & \mathcal{E}_{n}+\Delta_{e}\beta\\
\mathcal{E}_{n}-\Delta_{e}\beta & \Delta_{e}\beta\end{bmatrix}\begin{pmatrix}1\\
1\end{pmatrix}=\begin{pmatrix}\mathcal{E}_{n}\\
\mathcal{E}_{n}\end{pmatrix}\end{align}

For a Fermi velocity $v_{F}$, $E(k)-\mu\approx\hbar v_{F}(k-k_{F})$.
We can see that $u(k)$ is a Gaussian with width $\sqrt{\Delta_{e}/\hbar v_{F}}\approx\sqrt{\Delta_{0}/\hbar v_{F}\xi}$,
which justifies the substitution $k\rightarrow k_{F}$ earlier. %This implies that the wavefunctions for the original Hamiltonian \eqref{eq:sWaveVortexHam} has localization length of $\sqrt{\hbar v_{F}/\Delta_{e}}\approx\sqrt{\hbar v_{F}\xi/\Delta_{0}}$.

We would like to point the reader to one last detail before moving
on to the general solution. Putting our solution back in to the wavefunction
\eqref{eq:CircularWf} gives \begin{align}
\psi(k,\theta)=\frac{u(k)e^{in\theta}}{\sqrt{k}}\begin{pmatrix}e^{-i\theta}\\
-i\end{pmatrix}.\end{align}
 The pseudo-spinor (in the Nambu basis) is an eigenstate of $\mathbf{t}\cdot\bm{\Pi}$,
where $\mathbf{t}$ is a vector tangent to the FS at $(k_{F},\theta)$.
The pseudo-spin locking to the the momentum gives rise to the $\pi$
phase around the FS in the Hamiltonian \eqref{eq:CircularWHamWdag}
and Eq.~\eqref{eq:CircularVortexEnergies}.

\subsection{General solution without rotational symmetry}

The solution is similar in spirit to the case with circular symmetry.
Now, we label the momentum by energy contours $(E,\eta)$ instead
of $(r,\theta)$ and demand constant $E$ contours to be orthogonal
to constant $\eta$ contours. Similar to $\theta$, $\eta$ is periodic
with periodicity of $2\pi$.

The idea of the derivation is as follows. We rewrite the Hamiltonian
in a Jackiw-Rebbi form as a function $E$, perpendicular to the FS.
Let $\Gamma^{1},\Gamma^{2},\Gamma^{3}$ be matrices which anticommute
with each other, and square to the identity matrix. Then the differential
equation \begin{align}
\Gamma^{1}(i\Delta_{e}\partial_{E})+i\Gamma^{2}(E-\mu)+\Gamma^{3}f(\eta)\label{eq:JackiwRebbi}\end{align}
 has a bound state when $E-\mu$ changes sign, the bound state is
an eigenvector of $\Gamma^{3}$ and has energy $f(\eta)$~\cite{JackiwRebbi76}.
%We know there is a bound state whenever $E(\vr)$ changes sign.
The remaining $\eta$ degree of freedom governs the existence of zero
energy states, by requiring the wavefunction $\psi(E,\eta)$ to be
single valued.

\begin{figure}[tb]
\includegraphics[width=0.4\textwidth,keepaspectratio=true]{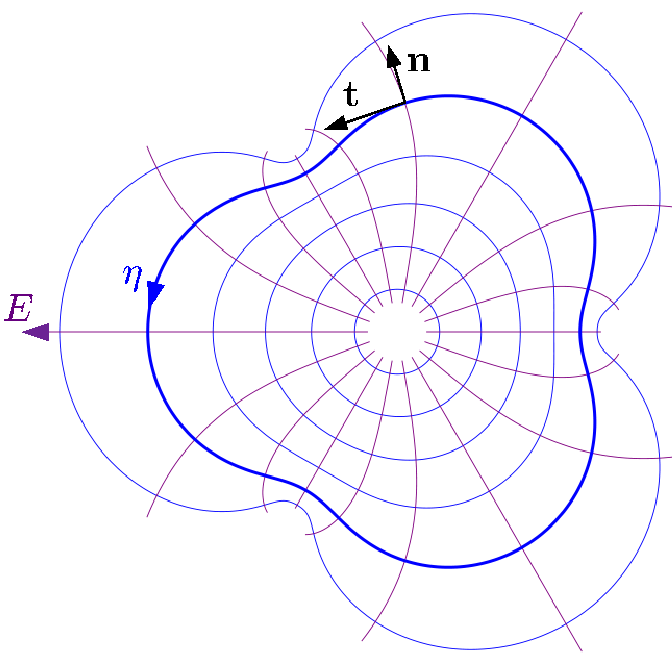} \caption{(Color online) Parameterization of the reciprocal space with orthogonal
coordinates $(E,\eta)$. The (blue) closed contours are plotted for
constant energies $E$, with the bold contour being the FS at the
chemical potential $\mu$. The (purple) open contours are at constant
$\eta$, running from $0$ to $2\pi$. At each point, $\mathbf{t}(E,\eta)=\frac{\partial\vk}{\partial\eta}$
and $\mathbf{n}(E,\eta)=\frac{\partial\vk}{\partial E}$ are vectors
tangent and normal, respectively, to the energy contours. Notice that,
for a closed FS, the direction of the vectors $\mathbf{t},\mathbf{n}$
rotate once counterclockwise as $\eta$ is varied around the surface. }

\label{fig:FermiSurfaceEeta} 
\end{figure}

Define the vectors tangent and normal to the energy contours $\mathbf{t}=\frac{\partial\vk}{\partial\eta}$,
$\mathbf{n}=\frac{\partial\vk}{\partial E}$. Clearly, $\mathbf{t}\perp\mathbf{n}$.
(Fig.~\ref{fig:FermiSurfaceEeta}). The derivatives and connections
in $(E,\eta)$ coordinates are related to those in the Cartesian coordinates:
\begin{equation}
\begin{pmatrix}\dE\\
\deta\end{pmatrix}=\begin{bmatrix}n^{x} & n^{y}\\
t^{x} & t^{y}\end{bmatrix}\begin{pmatrix}\partial_{k_{x}}\\
\partial_{k_{y}}\end{pmatrix},\end{equation}
 and we define the $2\times2$ matrix to be $J^{-1}$. \begin{align}
\begin{pmatrix}\partial_{k_{x}}\\
\partial_{k_{y}}\end{pmatrix} & =J\begin{pmatrix}\dE\\
\deta\end{pmatrix}, & \begin{pmatrix}A_{x}\\
A_{y}\end{pmatrix} & =J\begin{pmatrix}A_{E}\\
A_{\eta}\end{pmatrix},\end{align}
 where $J$ is the Jacobian matrix: \begin{align}
J=\frac{\partial(E,\eta)}{\partial(k_{x},k_{y})}=\begin{bmatrix}J_{x}^{E} & J_{x}^{\eta}\\
J_{y}^{E} & J_{y}^{\eta}\end{bmatrix}.\end{align}
 From $J^{-1}J=1$, we can see that $\JE\cdot\mathbf{t}=0$, $\Jeta\cdot\mathbf{n}=0$,
hence $\JE\parallel\mathbf{n}$, $\Jeta\parallel\mathbf{t}$.

First we rewrite the pairing term in terms of $E$ and $\eta$. $i(\partial_{k_{x}}-i\partial_{k_{y}})+A_{x}-iA_{y}=d_{E}+d_{\eta}$,
where \begin{subequations} \label{eq:defdEeta1} \begin{align}
d_{E} & =i(J_{x}^{E}-iJ_{y}^{E})(\dE-iA_{E}),\\
d_{\eta} & =i(J_{x}^{\eta}-iJ_{y}^{\eta})(\deta-iA_{\eta}).\end{align}
 \end{subequations} It is always possible to find a gauge transformation
which eliminates $A_{E}$ near the FS. The transformation is of the
form $U_{A}^{\dag}=\mathcal{P}\exp\big[i\int_{0}^{E}\! A_{E}(E')dE'\big]$,
where $\mathcal{P}$ is the path-ordering operator. Since $\dE U_{A}^{\dag}=iA_{E}U_{A}^{\dag}$,
\begin{align}
U_{A}(\dE-iA_{E})U_{A}^{\dag}=\dE\end{align}
 This transformation will alter $A_{\eta}$, since the derivative
$\deta$ acts on $U_{A}^{\dag}$. In general, it impossible to make
both $A_{E}$ and $A_{\eta}$ disappear.

We make the substitution $J_{x}^{E}-iJ_{y}^{E}=|\JE|e^{-i\thetan}$,
where $\thetan$ gives the direction of the normal vector $\mathbf{n}$.
As $\mathbf{t}$ is perpendicular to $\mathbf{n}$, $J_{x}^{\eta}-iJ_{y}^{\eta}=-i|\Jeta|e^{-i\thetan}$,
Eq.~\eqref{eq:defdEeta1} becomes \begin{subequations} \begin{align}
d_{E} & =ie^{-i\thetan}|\JE|\dE,\\
d_{\eta} & =e^{-i\thetan}|\Jeta|(\deta-iA_{\eta}).\end{align}
 \end{subequations} At the moment, our Hamiltonian \eqref{eq:p+ipHam}
is of the form: \begin{widetext} \begin{align}
\tilde{\mathcal{H}} & =\begin{bmatrix}E-\mu & \Delta_{e}e^{-i\thetan}\left(i|\JE|\dE+|\Jeta|(\deta-iA_{\eta})\right)\\
\Delta_{e}e^{i\thetan}\left(i|\JE|\dE-|\Jeta|(\deta-iA_{\eta})\right) & -E+\mu\end{bmatrix}.\label{eq:HamEeta}\end{align}
 \end{widetext}

The angle $\thetan$ rotates by $+2\pi$ around a closed FS as $\eta$
varies from $0$ to $2\pi$. The $e^{i\thetan}$ phase in the off
diagonal terms of the Hamiltonian give rise to a $\pi$ phase in the
eigenstates. We transform this phase away via the unitary transformation:
\begin{align}
U_{ph}=\begin{bmatrix}e^{i\thetan}\\
 & 1\end{bmatrix},\end{align}
 such that \begin{align}
U_{ph}\,\tilde{\mathcal{H}}\, U_{ph}^{\dag} & =\begin{bmatrix}E-\mu & \Delta_{e}D^{-}\\
\Delta_{e}D^{+} & \mu-E\end{bmatrix}.\end{align}
 where \begin{subequations} \begin{align}
D^{-} & =i|\JE|\dE-i|\Jeta|(i\deta+A_{\eta})\\
D^{+} & =i|\JE|\dE+i|\Jeta|(i\deta+A_{\eta})+i|\Jeta|\frac{\partial\thetan}{\partial\eta}\end{align}
 \end{subequations} We have ignored the term $\dE\thetan$ from the
assumption that the FS is smooth (no cusps). The term $\deta\thetan$
is extremely important as it will give us the $\pi$ Berry phase shift.

At any fixed value of $\eta$, we write the Hamiltonian in the form
of Eq.~\eqref{eq:JackiwRebbi}: %as a function of $E$ has mass term $E-\mu$ which changes sign \cite{JackiwRebbi76}.
\begin{align}
U_{ph}\tilde{\mathcal{H}}U_{ph}^{\dag} & =(E-\mu)\tau_{z}+\Delta_{e}D^{x}\tau_{x}+\Delta_{e}D^{y}\tau_{y},\label{eq:UHUdagJR}\end{align}
 where $D^{x}$ and $D^{y}$ are the symmetric and the antisymmetric
parts of off the diagonal elements $D^{\pm}$. \begin{subequations}
\begin{align}
D^{x} & =\frac{D^{+}+D^{-}}{2}=i|\JE|\dE+i\frac{|\Jeta|}{2}\frac{\partial\thetan}{\partial\eta},\label{eq:JRHamDx}\\
D^{y} & =\frac{D^{+}-D^{-}}{2i}=|\Jeta|\left(i\deta+A_{\eta}+\frac{1}{2}\frac{\partial\thetan}{\partial\eta}\right).\label{eq:JRHamDy}\end{align}
 \end{subequations} In the first expression \eqref{eq:JRHamDx},
there is an extra term $\frac{i|\Jeta|}{2}\frac{\partial\thetan}{\partial\eta}$,
which we can absorb in to the energy derivative via the transformation
\begin{align}
\exp\left[\int\! g_{\eta}\, dE\right]\,\dE\,\exp\left[-\int\! g_{\eta}\, dE\right]=\dE-g_{\eta}\end{align}
 where $g_{\eta}(E)=\frac{|\Jeta|}{2|\JE|}\frac{\partial\thetan}{\partial\eta}$.
This introduces a term $\deta g_{\eta}$ in $D^{y}$, but is irrelevant
as $g_{\eta}$ is single-valued. (In the case with rotational symmetry,
the term $\exp\int\! g_{\eta}$ is $k^{1/2}$ in \eqref{eq:CircularWph}.)

The ($E$ dependent portion of the) solution to Jackiw-Rebbi Hamiltonian
\eqref{eq:UHUdagJR} is: \begin{align}
\psi(E,\eta)\sim u(\eta)\exp\left[-\int\frac{E-\mu}{\Delta_{e}|\JE|}dE\right]\begin{pmatrix}1\\
-i\end{pmatrix},\end{align}
 by treating $D^{y}$ as a constant (independent of $E$) on the energy
scale of $\sqrt{\Delta_{e}|\JE|}$. The effective Hamiltonian is $\mathcal{H}_{\eta}=-\Delta_{e}D^{y}$.

%Finally, we solve for $\Delta_e D^y(\eta) u(\eta) = 0$ to compute the energy of the vortex bound states, where $u(\eta)$ is single-valued.
Finally, we solve for $D^{y}(\eta)u(\eta)=0$ for a Majorana vortex
bound state, subject to the constraint that $u(\eta)$ is single-valued.
We can solve for $u(\eta)$ explicitly: \begin{align}
u(\eta)=\mathcal{P}\exp\left[i\int_{0}^{\eta}\!\! d\eta'\left(A_{\eta}(\eta')+\frac{1}{2}\frac{\partial\thetan}{\partial\eta}\right)\right]u(0).\end{align}
 Since $\thetan$ winds by $2\pi$ around a \emph{closed} FS, $\oint\!\frac{1}{2}\deta\thetan=\pi$
gives an overall phase of$-1$. We have $u(2\pi)=-U_{B}u(0)$, where
the Berry phase of the FS is \begin{align}
U_{B}=\mathcal{P}\exp\left[i\oint_{0}^{2\pi}\!\! A_{\eta}d\eta\right].\end{align}
 Hence a solution exists for every $-1$ eigenvalue of $U_{B}$.

%As $\frac{1}{2}\deta\thetan$ integrates to $\pi$ and $A_\eta$ integrates to $\phi_F/2\pi$ around the FS, where $\phi_F$ is the Berry phase, a zero energy solution exist if and only if $\phi_F = \pi$.
%A zero energy solution exist if we can find a single-valued $u(\eta)$.

\subsection{Vortex bound states}

We can solve for the spectrum Caroli-de~Gennes-Matricon bound states~\cite{CdGMSCVortexStates64}.
For an arbitrary energy $\mathcal{E}$, the `angular' portion of the
wavefunction satisfies $-\Delta_{e}D^{y}u(\eta)=\mathcal{E}u(\eta)$,
equivalent to: \begin{align}
-i\deta u(\eta) & =\left(\frac{\mathcal{E}}{\Delta_{e}|\Jeta|}+A_{\eta}+\frac{1}{2}\frac{\partial\thetan}{\partial\eta}\right)_{E=\mu}u(\eta).\label{eq:etaDiffEq}\end{align}
 Note that $|\Jeta|^{-1}=|\mathbf{t}|$. The solution for $u(\eta)$
is a the path-ordered exponential \begin{widetext} \begin{align}
u(\eta) & =\mathcal{P}\exp\left[i\int_{0}^{\eta}\!\left(\frac{\mathcal{E}|\mathbf{t}|}{\Delta_{e}}+A_{\eta}+\frac{1}{2}\frac{\partial\thetan}{\partial\eta}\right)_{E=\mu}\! d\eta'\right]u(0).\label{eq:uetaSol}\end{align}
 The full solution to the Hamiltonian \eqref{eq:HamEeta} is: \begin{align}
\psi(E,\eta) & \;\propto\; u(\eta)\exp\left[-\int^{E}\!\left(\frac{E'-\mu}{\Delta_{e}|\JE(E')|}+g_{\eta}(E')\right)dE'\right]\begin{pmatrix}e^{-i\thetan}\\
-i\end{pmatrix}.\end{align}
 \end{widetext} %where $u(\eta)$ is given by \eqref{eq:uetaSol}.

While $A_{\eta}$ is a hermitian matrix, $\mathcal{E}|\mathbf{t}|/\Delta_{e}$
and $\deta\thetan$ are simply numbers, hence the integral may be
evaluated separately for each term. The integral $\oint_{0}^{2\pi}\!|\mathbf{t}|d\eta$
is simply the perimeter of the FS $l_{F}$. The integral $\frac{1}{2}\oint_{0}^{2\pi}\!\deta\thetan d\eta$
is $\pi$ for a closed FS, and $0$ for an open FS (modulo $2\pi$).
The single-valued requirement $u(2\pi)=u(0)$ (for a closed FS) becomes:
\begin{align}
u(0)=-\exp\left(i\frac{l_{F}}{\Delta_{e}}\mathcal{E}\right)U_{B}u(0).\end{align}

%Notice that the spinor $(e^{-i\thetan}, -i)^T$ is locked to the tangent vector along the FS, required by the Jackiw-Rebbi Hamiltonian.  Just as in the rotationally symmetric case, a $-1$ phase shift exists traversing around a closed FS.

%Also note that $\int\! |\Jeta|^{-1} d\eta = \int\! |\mathbf{t}|d\eta$ gives the path length along the constant $E$ contour, and so the loop integral $l_F = \oint\! \frac{d\eta}{|\Jeta|}$ is the perimeter of the FS.
From this, we can calculate the allowed energies for an arbitrary
Berry phase: \begin{align}
\mathcal{E}_{n}=\frac{\Delta_{e}}{l_{F}}(2\pi n+\pi-\phi),\end{align}
 for integers $n$, where $e^{i\phi}$ are the eigenvalues of $U_{B}$.
Note that the eigenvalues always come in pairs $\pm\phi$ due to the
particle-hole symmetry of the system (considering both $k_{z}$ and
$-k_{z}$ slices of the BZ).

%%%%%%%%%%%%%%%%%%%%%%%%%%%%%%%%%%%%%%%%%%%%%%%%%%%%%%%%%%%%%%%%%%%%%%%%%%%%%%%%%%%%%%%%%%%%%%%%%%%%

\subsection{Numerical calculation of the Berry phase for Bi$_{2}$Se$_{3}$}

The insulating phase of Bi$_{2}$Se$_{3}$ is a strong TI, with a
band inversion occurring at the $\Gamma$ point in its rhombohedral
Brillouin zone. Cu$_{x}$Bi$_{2}$Se$_{3}$ is reported to superconduct
below $T_{c}=\unit[3.8]{K}$~\cite{SuperconductingCuxBi2Se3,SCdoping}.
Before the superconducting transition, the carrier (electron) density
is approximately $\unit[2\times10^{20}]{cm^{-3}}$ from Hall measurements~\cite{SCdoping}.
Using the effective eight-band model from Ref.~\cite{LiuModelHamiltonian},
we estimate $\mu$ theoretically in this material from the carrier
density to be $\approx\unit[0.4]{eV}$ relative to the conduction
band bottom. However, photoemission measurements show $\mu\approx\unit[0.25]{eV}$
above the bottom of the conduction band at the optimal doping ($x=0.12$)~\cite{CuxBi2Se3ARPES}.

\begin{figure}[hb]

\begin{centering}
\includegraphics[width=0.8\columnwidth,keepaspectratio=true]{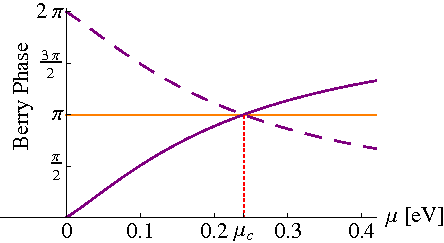} 
\par\end{centering}

\caption{Berry phase eigenvalues $\phi_{B}$ for Bi$_{2}$Se$_{3}$ as a function
of $\mu$ relative to the bottom of the conduction band. The eigenvalues
appear in $\pm$ pairs and are only defined modulo $2\pi$. The phases
$\phi_{B}$ are zero (modulo $2\pi$) at both the conduction band
minimum and at energies far above the conduction band. Clearly, the
$\phi_{B}=\pi$ at $\mu_{c}=\unit[0.24]{eV}$, signalling a VPT. Photoemission
measurements show $\mu\approx\unit[0.25]{eV}\sim\mu_{c}$ above the
conduction band minimum at optimal doping~\cite{CuxBi2Se3ARPES},
and hence Cu doped Bi$_{2}$Se$_{3}$ is predicted to be near the
vortex phase transition. }

\label{fig:Berry-phase-plot}
\end{figure}

The conduction band minimum is at the $\Gamma$-point. For small carrier
densities, we expect the FS to be centered around $\Gamma$. Using
the same model from Ref.~\cite{LiuModelHamiltonian}, we determine
$\mu_{c}$ for this material by calculating the Berry phase eigenvalues
for a FS around the $\Gamma$ point numerically as a function of $\mu$
(See Fig.~\ref{fig:Berry-phase-plot}). %This procedure gives $\mu_{c}\approx0.24eV$
%above the conduction band minimum for a vortex along the $c$-axis
%of the crystal.
%$\mu>\mu_{c}$,
%indicates c-axis vortices are topologically trivial.

This calculation is done by discretizing the FS contour and making
the pair of Bloch functions continuous and the Berry connection vanish
at all but one points on this contour. More precisely, we parameterize
the FS via $\eta\in[0,2\pi]$ and compute the eigenstates $\ket{\varphi^{\nu}(\eta)}$
for $\nu=1,2$. The phase is chosen such that $A_{\eta}^{\mu\nu}=i\bra{\varphi^{\mu}(\eta)}\deta\ket{\varphi^{\nu}(\eta)}=0$
for $0<\eta<2\pi$ along the FS.

In general, the Bloch functions will not be single-valued, \textit{i.e.}
$\ket{\varphi^{\nu}(2\pi)}\neq\ket{\varphi^{\nu}(0)}$. The unitary
transformation required to rectify the discontinuity at this one point
is precisely the non-Abelian phase: $\ket{\varphi^{\mu}(0)}=[U_{B}]^{\mu\nu}\ket{\varphi^{\nu}(2\pi)}$.
Because of time-reversal symmetry in the normal phase of Bi$_{2}$Se$_{3}$,
the Berry phase $U_{B}\in SU(2)$ has eigenvalues $e^{\pm i\phi}$
which come in complex conjugate pairs. Fig.~\ref{fig:Berry-phase-plot}
shows the variation of the Berry phase eigenvalues $\phi$ as a function
of $\mu$ for the FS of Bi$_{2}$Se$_{3}$ in the $k_{c}=0$ plane,
where the c-axis is normal to the layers. Here, $\mu$ is measured
relative to the bottom of the conduction band at the $\Gamma$ point,
which is where the band inversion occurs. Hence $\mu_{c}\approx\unit[0.24]{eV}$,
which happens to be close to the chemical potential for these doped
materials according to ARPES. %we expect the vortex core of Cu doped Bi$_{2}$Se$_{3}$ to be topologically trivial; there are no surface MZMs.
%Clearly, the eigenvalues hit $\pi$ at $\mu_{c}\approx0.24eV$.

%

\bibliography{FullDraft}